\documentclass[12pt]{iopart}

\usepackage{subfigure}
\usepackage{graphicx}
\usepackage{threeparttable}
\usepackage{cite}
\usepackage{booktabs}
\usepackage{dcolumn}
\usepackage{bm}
\usepackage{longtable}
\graphicspath{ {FIGURE/} }
\usepackage{comment}
\usepackage{fancyhdr}

\fancypagestyle{plain}{
\fancyhead{}
 
}

\usepackage{color}
\usepackage{soul}
 
\usepackage{iopams}

\begin{document}

\title{Extreme Gravitational Interactions in the Problem of Three Black Holes in General Relativity}

\author{M. Imbrogno$^1$, C. Meringolo$^1$ and S. Servidio$^1$}

\address{$^1$Dipartimento di Fisica, Universit\`a della Calabria, I-87036 Cosenza, Italy}
\ead{sergio.servidio@fis.unical.it}
\vspace{10pt}

\begin{abstract}
We study the three-body problem going from Newtonian mechanics to general relativity. In the classical case, we model the interactions in a typical chaotic configuration, identifying Extreme Gravitational Interactions (EGIs), namely transients in which the system manifests complex, highly-energetic dynamics. We then concentrate on the main part of the work, by selecting these EGIs as initial data for the general relativistic case, and performing a campaign of numerical relativity simulations. To provide a comprehensive menu of cases, we investigate different global configurations. By comparing with the more ``quiet'' two-body inspiral, we observe strong nonlinear emission of gravitational waves. The multi-body signals have been inspected by employing both Fourier and wavelet analyses, showing net differences among the global configurations. The wavelet analysis reveals the reminiscence of the EGIs in the three black holes problem. Such a survey of simulations might be a guide for future observations.

\vspace{15pt}
\end{abstract}

\section{Introduction}
The gravitational three-body problem represents the non-linear problem {\it par excellence} \cite{mj2006three}.  There are several applications of this problem, going from microphysics to large-scale astrophysics. In this cross-scale variety of examples, the problem remains puzzling, despite years of studies, since the bodies have an erratic behavior (see \cite{valtonen1996triple} for a review). From the classical perspective, for example, in problems of celestial mechanics, one may almost always identify a binary and a third body. A binary can be treated as a single entity with certain internal properties (like a molecule). The couple interacts with a third body, once or more frequently, with a resultant change in the internal properties of the binary. During the interaction, the system conserves the total energy, the mass, the momentum, and the angular momentum, even though the trajectories of the single bodies are unpredictable \cite{monaghan1976statistical, mj2006three}. During the complex dynamics, the system redistributes energy and angular momentum efficiently; as three-body scattering goes on, the population may gain speed and become ‘heated’. This has profound consequences on the structure and evolution of clusters. Three- or many-body interactions are therefore expected to be common in clusters and galactic cores \cite{Miller2002four}. Multiple  black holes might also be formed in galactic nuclei undergoing sequential mergers \cite{makino1990bottlenecks, valtonen1996triple}, triple quasar systems \cite{djorgovski2007discovery}, globular clusters, and galactic disks \cite{gultekin2003three, berukoff2006cluster, gualandris2005three}.

There is a good chance of having, at some point in the interaction, the three bodies strongly dominated by their mutual gravitational attraction. In this regime, when masses are large, the chaotic three-body dynamics should be treated via general relativistic models, and, when distances are on the order of the gravitational radii, merging events are likely to occur \cite{antonini2014black, samsing2014formation}. In globular clusters, indeed, $N$-body interactions are important in the formation of massive black holes \cite{gultekin2003three, miller2019new, gultekin2004growth}. In such a system, the conservation of the classic invariants needs to be revised in terms of general relativity (GR) \cite{galaviz2010numerical, Lousto2008foundations}. 

The mergers of compact massive objects represent one of the most suggestive and extraordinary events in nature. According to general relativity, these systems are powerful sources of gravitational waves -- disturbances of the spacetime. However, the gravitational radiation, detected through modern experimental settings, cannot unambiguously characterize what happens in the vicinity of the merging region, and complementary studies of numerical relativity are needed \cite{Pretorius2005evolution, Pretorius2006simulation, ceverino2015early, bournaud2008high, lehner2014numerical}. It is, therefore, crucial to construct guidance for future observational campaigns, by exploring a wide class of possibilities for the three-body configurations, by varying for example parameters such as their initial distribution, their initial global angular momentum, their spin, and their masses. The understanding of such extreme, nonlinear interactions represents a new challenge for the comprehension of cosmological problems and needs to be tackled via appropriate models, following the Einstein's vision of gravitation \cite{Campanelli2008close, silsbee2017lidov}.

In this work, we investigate the main features of the three-body problem, going from classical mechanics to GR. We start with the study of the behavior of three point-like bodies according to classical Newton's laws. By using a Lagrangian numerical model, we initially consider a geometrical distribution, known as the \textit{Burrau problem} \cite{Lousto2008foundations}. Subsequently, we study the same problem by varying the global momentum and the masses, to build a carpet for the study of the analogous dynamics in the case of highly-massive objects. In the second part, we perform direct numerical simulations of GR, by using as initial data the conditions coming from the Newtonian case. We finally inspect the waveform of the gravitational radiation emitted from the system, comparing it with a two-body inspiral. These studies might be useful for the description and understanding of gravitational signals \cite{danzmann1996lisa, armano2009lisa, baker2021high}.

The paper is organized as follows. In Section \ref{sec:newtn}, we summarize the classical results of the chaotic motion of the three-body problem, by solving numerically the equation of motion and by introducing measurements of extreme body interactions. In Section \ref{sec:gr}, we tackle the problem from the point of view of general relativity, by introducing a $3+1$ formalism to integrate the Einstein field equations, presenting the Spectral-FIltered Numerical Gravity codE (\texttt{SFINGE}). Simulations of multi-black hole dynamics are presented in Section \ref{sec:inspir}, comparing the outcome of two-black-hole (2-BHs for brevity) inspiralling with the three-black hole dynamics initiated via the Newtonian code. Here we highlight the main differences via different spectral analyses of the produced signal, away from the sources. Finally, conclusions are discussed in Section \ref{sec:concl}.

\section{Three-bodies in the Newtonian case}
\label{sec:newtn}

\begin{table}[t]
\centering
\caption{\label{TABLE1}Initial parameters of the classical simulations in the four cases investigated. The values are reported in geometrical units, so the initial positions and velocities are normalized to the total mass of the system $M$.}

\scalebox{0.95}[0.95]{
\begin{tabular}{ccccc}
\br
Run & a & b & c & d \\
Name & Classical Burrau & Equal masses & Normalized masses & Spinning Burrau \\
\mr
$M_1$ & 3.0 & 0.33 & 0.25 & 0.25  \\
$M_2$ & 4.0 & 0.33 & 0.33 & 0.33  \\
$M_3$ & 5.0 & 0.33 & 0.42 & 0.42  \\
$\bm{r}_1/M$ & (1, 3, 0) & (1, 3, 0) & (1, 3, 0) & (1, 3, 0) \\
$\bm{r}_2/M$ & (-2, -1, 0) & (-2, -1, 0) & (-2, -1, 0) & (-2, -1, 0) \\
$\bm{r}_3/M$ & (1, -1, 0) & (1, -1, 0) & (1, -1, 0) & (1, -1, 0) \\
$\bm{v}_1/M$ & (0, 0, 0) & (0, 0, 0) & (0, 0, 0) & (-0.299, 0.050, 0) \\
$\bm{v}_2/M$ & (0, 0, 0) & (0, 0, 0) & (0, 0, 0) & (0.149, -0.100, 0) \\
$\bm{v}_3/M$ & (0, 0, 0) & (0, 0, 0) & (0, 0, 0) & (0.059, 0.050, 0) \\
\br
\end{tabular}
}
\end{table}

Despite the success in the understanding of simple (reduced) cases \cite{mj2006three, goldstein1980classical}, the solution to the general three-body problem remained elusive for about 200 years after the publication of Newton's \textit{Principia}. In the general three-body problem, all three masses are all of the same order and their initial positions and velocities are not arranged in any particular way. The difficulty of such a problem derives from the fact that there are no coordinate transformations that simplify the problem. This is in contrast to the two-body case where the solutions are found most easily in the center-of-mass coordinate system. In the three-body case, such transformation does not alleviate the problem, a difficulty that made the system analytically rather intractable, up to the modern age of computers. The numerical approach, indeed, revealed that the orbits are good examples of chaos in nature \cite{mardling2008resonance, liao2014physical, boyd1993chaotic}. 

The numerical three-body problem can be solved in different ways, although we do not concentrate our work on the classical integration. The problem has been solved several times with a variety of techniques \cite{aarseth1994close}, for example, with arbitrary precision regularization \cite{zwart2018numerical}. There is a large literature in this particular field, where other interesting approaches are used \cite{lehto2008mapping, Hernandez2020Are}, but the investigation of the classical case is out of the purpose of the present paper. As follows, we go through the main numerical results of the classical problem.

\begin{figure}
\centering
\hspace{1pt}  
\includegraphics[height=130mm,width=140mm]{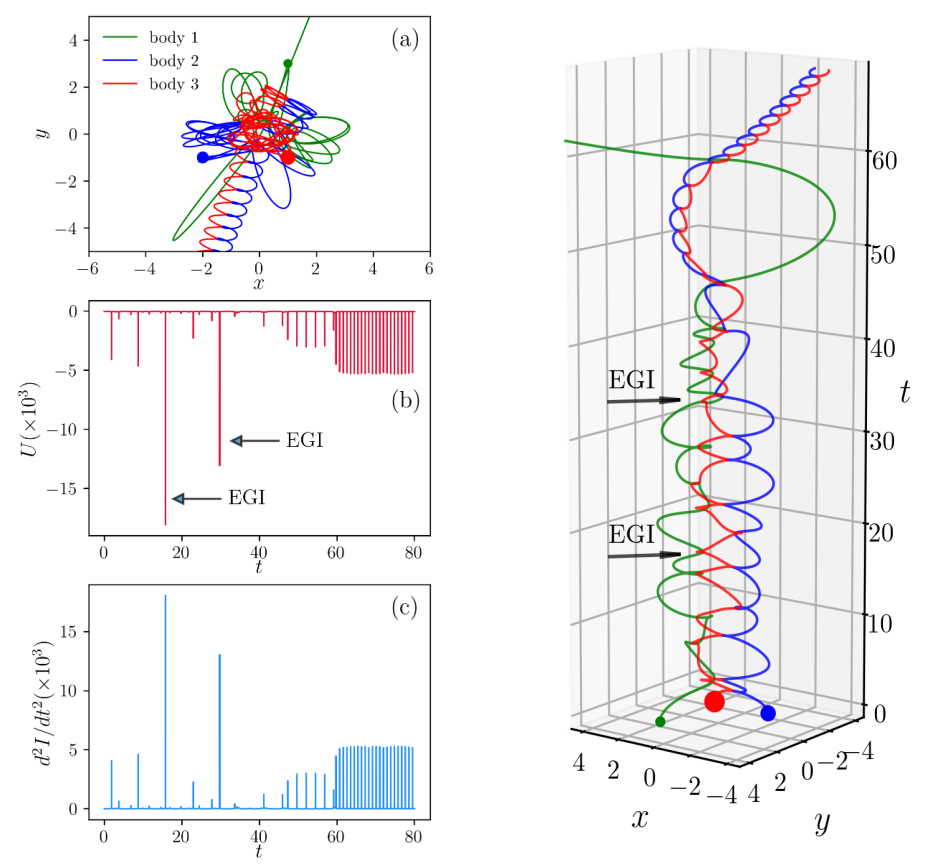}
\caption[ ]{\footnotesize{} Overview of the Burrau three-body problem in the classical case. Left: a) Trajectories starting from the Burrau configuration (the initial position of the bodies is emphasized by the colored balls), up to $t = 70$. (b) Time history of the potential energy. (c) Second time-derivative of the momentum of inertia for the system. Spikes in (b) and (c) correspond to typical EGIs. Right: time evolution of the trajectories of the three bodies starting from the Burrau configuration. One can easily distinguish the times at which the objects are very close to each other and interact strongly, i.e. the EGIs.}
\label{Figura1}
\end{figure}

The system of equations describing the multi-body system of $N$ point-like masses $M_i$, affected only by their mutual gravitational forces, is given by: 
\begin{eqnarray}
   && \dot{\bm{x}_i} = \bm{v}_i, \\
   && M_i \dot{\bm{v}_i} = - \sum_{j \ne i}^N \frac{M_i M_j}{(\bm{r}_i - \bm{r}_j)^2} \hat{\bm{r}}_{ij} = - \sum_{j \ne i}^N M_i M_j \frac{\bm{r}_i - \bm{r}_j}{|\bm{r}_i - \bm{r}_j|^3},
\end{eqnarray}
where we have used the geometrical units $c = G = 1$, $\bm{r}_i$ is the position of the $i^{th}$ body, $\bm{v}_i$ is its velocity, and $N = 3$ represents the number of bodies. We numerically solved the above equations, by using a fourth-order Runge-Kutta technique, with quadruple precision. We start with a particular initial configuration of the three-body problem that immediately leads to a chaotic motion, namely the so-called \textit{Pythagorean problem} proposed by Burrau in the early 1900s \cite{burrau1913numerische}. The point-like bodies, whose masses are $M_1= 3$, $M_2 = 4$ and $M_3 = 5$, are initially at the corners of a Pythagorean triangle. These are located essentially on a $xy$-plane ($z = 0$), as reported in Fig.~\ref{Figura1} (a) with (colored) bullets of mass-proportional size.

In the beginning, the bodies are all at rest. Burrau’s calculation revealed the typical behavior of a three-body system: two bodies approach each other have a close encounter, and then recede again. Subsequently, other two-body encounters were calculated by Burrau until he came to the end of his calculating capacity. Later work has shown that the solution to the problem is quite typical of initially bound three-body systems.  After many close two-body approaches a configuration arises which leads to an escape of one body and the formation of a binary by the other two bodies \cite{mj2006three}, as represented in Fig.~\ref{Figura1} (a), where we report the trajectories of the three bodies up to $t = 70$ for the chosen configuration.

We measured  the kinetic $T$, the potential $U$, and the total energy $E$ of the system, as a function of time, until one of the bodies separates from the remaining binary, which happens at about $t \approx 60$. The potential energy, representative of the gravitational interaction, is reported in Fig.~\ref{Figura1} (b). These interactions are extraordinarily intermittent. We call these bursts Extreme Gravitational Interactions (EGIs), where two (or all) of the three bodies are particularly close and the system has very large potential (and kinetic) energy.  At later times, one can note that, for $t \gtrsim 60$, we have small, periodic energy spikes.  This is because, at about this time, the system is split into a (regular) binary system and a single-motion body, so we lose the stochastic behavior typical of three bodies systems. The phenomenon of the ``escaper'' is largely investigated in the literature \cite{mj2006three}. To characterize the system configuration, especially in the most energetic periods, we measured the moment of inertia $I = \sum_i M_i r_i^2$ and its acceleration
\begin{equation}
     \ddot{I} = \frac{d^2I}{dt^2} = 2 M_1 \left( v_1^2 + \mathbf{r}_1 \cdot \ddot{\mathbf{r}}_1 \right) + 2 M_2 \left( v_2^2 + \mathbf{r}_2 \cdot \ddot{\mathbf{r}}_2 \right) + 2 M_3 \left( v_3^2 + \mathbf{r}_3 \cdot \ddot{\mathbf{r}}_3 \right).
    \end{equation}

As reported in Fig.~\ref{Figura1} (c), $\ddot{I}$ peaks exactly at the EGIs, as expected. Analogously, differentiating the moment of inertia twice with respect to the time one obtains the Lagrange-Jacobi identity $\ddot{I} = 4 T + 2 U = 4 E_0 + 2 |U|$ ($E_0$ is the total energy of the system) and a measure of the compactness of the three-body system: in coalescence events, $\ddot{I} \sim 2|U|$ and therefore it becomes very large.

To explore the parameter space, we performed a simulation campaign varying with respect to the Burrau case (Run a) the masses and the overall angular momentum (and hence velocities of the single bodies). The cases are summarized in Table \ref{TABLE1}. In the case with equal masses, Run b, we essentially have the same Burrau configuration as in Run a, but with total mass being unity. Similarly, in Run c, we have the same conditions as in Run a, but re-scaling the masses such that again the total mass in the system is one. Finally, in the ``Spinning Burrau case'' (Run d), we choose non-zero initial velocities, such that the total angular momentum remains conserved. Here the whole system has a global spin, which somehow confines the masses to interact in a limited region of space. This case might be of interest for modeling global galaxies coalescence.

\begin{figure}[t]
\centering 
\includegraphics[height=100mm,width=121mm]{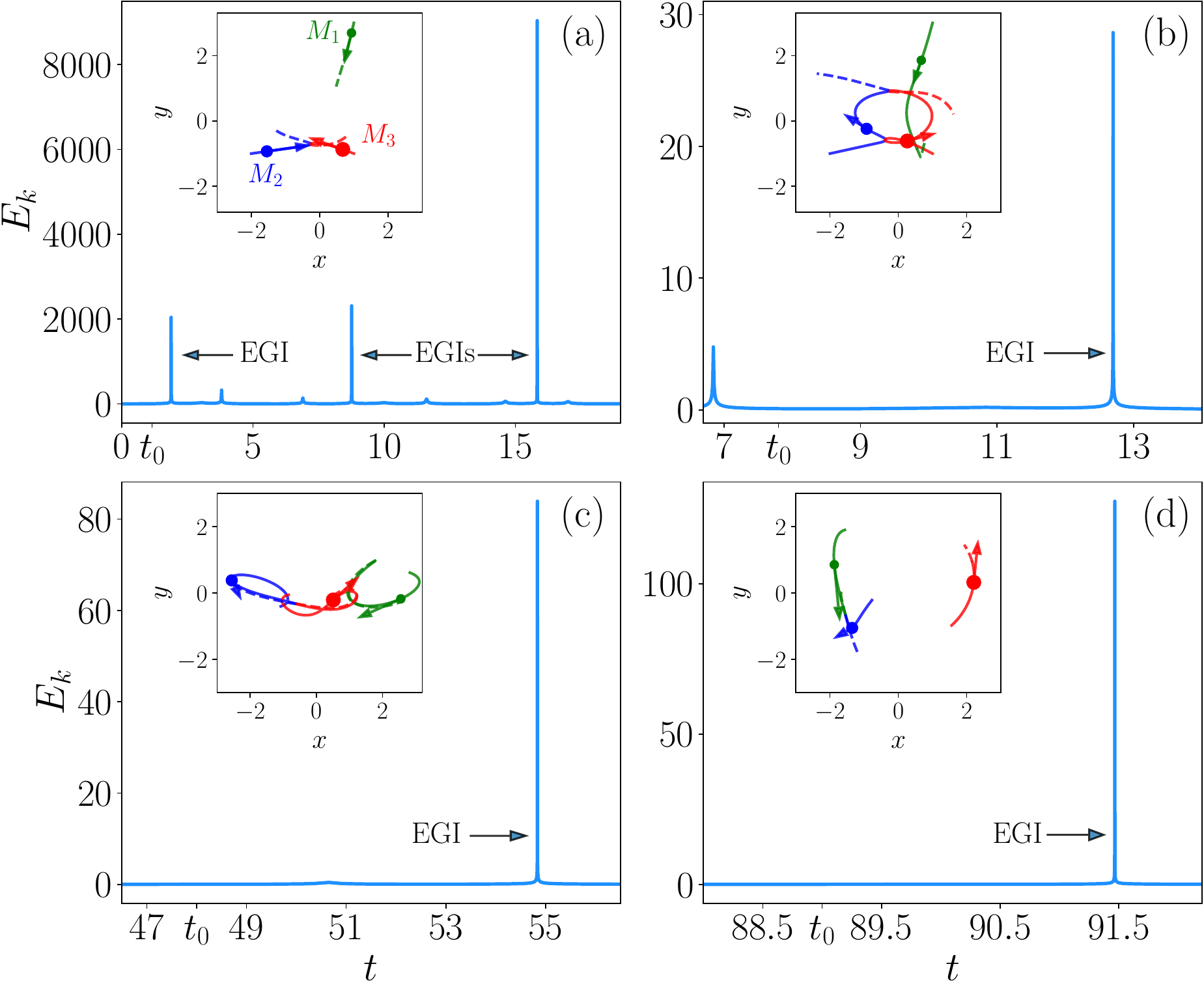}
\caption[ ]{\footnotesize{} Overview of the kinetic energy time history for each of the classic cases, namely, Run a--d in Table \ref{TABLE1} (each label corresponds to a configuration). In the insets, we show the position of the bodies, with mass-proportional bullets size, and the direction of the velocity vectors at the pre-EGI time $t_0$. This time has been highlighted with an arrow.}
\label{Classical_cases}
\end{figure}

In Fig.~\ref{Classical_cases} we present an overview of all the Newtonian simulations summarized in Table \ref{TABLE1}. In each panel, we represent the total kinetic energy of the system, concentrating therefore on time windows near extreme nonlinear interactions instants. In the inset of each panel, we report the position of the three bodies at a time that precedes an EGI. We labeled such time as $t_0$ and it will be of practical interest for our GR experiments (see later). Besides, we also report mass-proportional bullets, with the respective velocity arrows of the bodies, to give a global overview of each EGI event. The systems are all long-living and chaotic.

These numerical simulations might be of interest for celestial bodies with small masses, but they are unrealistic when it comes to massive bodies. In particular, it is natural to ask what happens when three large compact objects collide during an EGI. To answer such a question, one has to tackle the problem from a completely different perspective, invoking GR. We use the Newtonian configurations at the quiet times before EGIs, namely at $t_0$, as initial data for numerical relativity, exploring therefore the spacetime dynamics of Run a--d. In the next section, we provide the theoretical and numerical background for the GR investigation.

\section{Numerical relativity}
\label{sec:gr}

We started by performing Newtonian integration of a three-body configuration, by using rather different, general initial conditions, but just to extract the information for the general relativistic case. 

We now move toward the main goal of the work, namely to understand how Newtonian EGIs behave in the dynamical evolution
of this system, by investigating the possibility of observable gravitational wave signals.

The loss of energy and angular momentum from the system via gravitational radiation changes the basic constants of the three-body problem, and general relativity effects need to be incorporated. Such radiation leads to a decay of a binary orbit, and finally to a collapse of the binary black hole system into a single black hole. The final decay stages are very rapid and this might dramatically change the evolution of the three-body problem in the case of singular spacetimes.  What happens to three black holes when they strongly interact? The final result might be the escape of one of the black holes and the recoil of the binary in the opposite direction, or it can be the subsequent merger of the three massive objects \cite{Lousto2008foundations, Campanelli2008close, silsbee2017lidov}. We will focus on the interaction of the three black holes to extract the gravitational waves generated due to the loss of energy and angular momentum from the system, by varying also the initial configuration, as described in Table \ref{TABLE1}. To understand such emission, we will compare the two-body inspiralling with the three-body case.

We now introduce numerical techniques adequate for the stable and accurate evolution of multiple black-hole spacetimes. The technique is based on the \textit{moving puncture} approach \cite{Campanelli2006accurate}, treated by a novel pseudo-spectral method \cite{Meringolo2021spectral}. The system will be evolved on the basis of the Baumgarte-Shapiro-Shibata-Nakamura (BSSN) formulation of Einstein’s equations \cite{Alcubierre2008introduction, Baumgarte1998numerical, Shibata1995evolution}. 

The model is built starting from a conformal re-scaling of the three-metric
\begin{eqnarray}
    \gamma_{ij} = \psi^{4} \widetilde{\gamma}_{ij}, 
    \label{deco}
\end{eqnarray}
where $\psi$ constitutes the conformal factor. However, when considering singular spacetimes, it has been suggested that stable numerical solutions can be achieved by substituting $\chi = \psi^{-4}$ \cite{Campanelli2006accurate}. Reformulating the BSSN model in the key of the previous change of variable, the full set of equations can be written as

\begin{eqnarray}
 \label{evo1} 
 \partial_t \widetilde{\gamma}_{ij} = - 2 \alpha \widetilde{A}_{ij} + \beta^k \partial_k \widetilde{\gamma}_{ij} + \widetilde{\gamma}_{ik}\partial_j \beta^k + \widetilde{\gamma}_{kj}\partial_i \beta^k  -\frac{2}{3}\widetilde{\gamma}_{ij}\partial_k \beta^k, \\
 \label{evo2}
 \partial_t \chi = \frac{2}{3} \chi \left( \alpha K - \partial_i \beta^i \right) + \beta^i \partial_i \chi,\\
  \label{evo3}
 \partial_t   K = - D^2 \alpha + \alpha \Big( \widetilde{A}_{lm}\widetilde{A}^{lm} + \frac{1}{3} K^2 \Big) + \beta^i \partial_i K, \\
 \nonumber
 \partial_t\widetilde{A}_{ij}\!=\! \alpha \Big(\!K\widetilde{A}_{ij} - 2 \widetilde{A}_{ik} \widetilde{\gamma}^{m k} \widetilde{A}_{m j}\!\Big) + \chi\Big[\!\!-D_i D_j \alpha + \alpha R_{ij}
 \Big]^{TF} 
 \\
 ~~~~~~~~~~~~~
 \label{evo4}
 + \beta^k \partial_k \widetilde{A}_{ij} + \widetilde{A}_{ik}\partial_j \beta^k + \widetilde{A}_{kj}\partial_i \beta^k  -\frac{2}{3}\widetilde{A}_{ij}\partial_k \beta^k, \\
 \nonumber
 \label{evo5}
 \partial_t \widetilde{\Gamma}^i =  \widetilde{\gamma}^{lm} \partial_l \partial_m \beta^i + \frac{1}{3} \widetilde{\gamma}^{il} \partial_l \partial_m \beta^m  +  \beta^k \partial_k \widetilde{\Gamma}^i - \widetilde{\Gamma}^k \partial_k \beta^i + \frac{2}{3} \widetilde{\Gamma}^i \partial_k \beta^k 
 \\
 ~~~~~~~~~~~~~
 -  2 \widetilde{A}^{ik} \partial_k \alpha
 + \alpha \Big(  2\widetilde{\Gamma}^i_{lm} \widetilde{A}^{lm}
 - \frac{3}{\chi}\widetilde{A}^{ik} \partial_k \chi 
 - \frac{4}{3} \widetilde{\gamma}^{ik} \partial_k K  \Big),  \\
  \label{slicing1}
\partial_t \alpha = - \, \alpha^2 f(\alpha) K + \beta^i \partial_i \alpha, \\ 
  \label{slicing2} 
\partial_t \beta^i = \frac{3}{4} B^i , \\
  \label{slicing3}
\partial_t  B^i = \partial_t \widetilde{\Gamma}^i - \beta^j \partial_j  \widetilde{\Gamma}^i - \eta B^i.
\end{eqnarray}
In the above equations, the extrinsic curvature $K_{ij}$ is decomposed in the trace $K$ and its trace-free part $A_{ij}$. The latter, similarly to the conformal transformation of the metric, is expressed as $A_{ij}=\chi^{-1}\widetilde{A}_{ij}$, where $\widetilde{A}_{ij} = \chi \big( K_{ij} - \frac{1}{3} \gamma_{ij} K \big)$. Additionally, as can be noted, this approach involves the introduction of a new field, namely the contracted Christoffel symbols related to the conformal spatial metric  $\widetilde{\Gamma}^i = \widetilde{\gamma}^{jk} \widetilde{\Gamma}^i_{jk}$, which evolves separately according to equation (\ref{evo5}). The lapse function is represented by $\alpha$, $D_i$ is the covariant derivative in the physical space, $D^2 = \gamma^{ij} D_i D_j$, and $\beta^k$ is the typical shift vector of ADM slicing. The superscript ``$TF$'' stands for the trace-free part of a tensor.

The last part of the above system, namely the equations (\ref{slicing1})-(\ref{slicing3}), represents the evolution of the lapse and the shift vector, according  to the Bona-Massó family of slicing conditions \cite{Bona1989einstein, Bona1995new,Baumgarte1998numerical}. As in the existing literature \cite{Campanelli2006accurate}, $\eta$ is a positive constant (here $\eta = 2.8$) and the factor $3/4$ is an arbitrary coefficient that leads to stable numerical results \cite{Alcubierre2003towards}. These choices are known as the  ``Gamma-driver'' shift condition, which has been found to be extremely robust with puncture initial data, checking both the slice stretching and the shear due to the rotation of the black holes \cite{Alcubierre2008introduction}. Last but not least, $f( \alpha )$ is an arbitrary function that is set to $f( \alpha ) = 1/\alpha$ and corresponds to the ``$1 + \log$'' slicing conditions -- a typical choice in black holes simulations.

\begin{table}[t]
\centering
\caption{\label{TABLE2} Parameters of the BSSN simulations, based on the Newtonian cases in Table \ref{TABLE1}, at a given time $t_0$ before EGIs. The values are reported in geometrical units. The last line indicates the initial Arnowitt-Deser-Misner mass, evaluated as suggested by Baumgarte \cite{baumgarte2000innermost}.}

\scalebox{0.678}[0.728]{
\begin{tabular}{ccccccc}
\br
RUN & I & II & III & IV & V & VI \\
Name & 2 BHs & Classical Burrau & Equal masses & Normalized masses & Spinning Burrau & Intrinsic Spin \\
\mr
$M_1$ & 0.450 & 0.250 & 0.333 & 0.250 & 0.250 & 0.250 \\
$M_2$ & 0.450 & 0.333 & 0.333 & 0.333  & 0.333 & 0.333 \\
$M_3$ & // & 0.417 & 0.333 & 0.417 & 0.417 & 0.417 \\
$\bm{r}_1/M$ & (0, 1.151, 0) & (0.931, 2.695, 0) & (0.671, 1.857, 0) & (2.554, -0.172, 0) & (-1.874, 0.856, 0) & (-1.874, 0.856, 0) \\
$\bm{r}_2/M$ & (0, -1.151, 0.0) & (-1.544, -0.931, 0) & (-0.929, -0.242 , 0) & (-2.557, 0.385, 0) & (-1.352, -1.048, 0) & (-1.352, -1.048, 0) \\
$\bm{r}_3/M$ & // & (0.676, -0.872, 0) & (0.259, -0.615, 0) & (0.513, -0.205, 0) & (2.206, 0.325, 0) & (2.206, 0.325, 0) \\
$\bm{v}_1/M$ & (0.335, 0, 0) & (-0.037, -0.039, 0) & (-0.119, -0.358, 0) & (-0.296, -0.132, 0) & (0.036, -0.382, 0) & (0.036, -0.382, 0) \\
$\bm{v}_2/M$ & (-0.335, 0, 0) & (0.251, 0.037, 0) & (-0.336, 0.229, 0) & (0.034, -0.070, 0) & (-0.059, -0.035, 0) & (-0.059, -0.035, 0) \\
$\bm{v}_3/M$ & // & (-0.179, 0.066, 0) & (0.455, 0.129, 0) & (0.150, 0.135, 0) & (0.025, 0.258, 0) & (0.025, 0.258, 0) \\
$\bm{J}_1/M^2$ & (0, 0, 0) & (0, 0, 0)  & (0, 0, 0) & (0, 0, 0) & (0, 0, 0) & (0, 0, 0.100) \\
$\bm{J}_2/M^2$ & (0, 0, 0)  & (0, 0, 0)  & (0, 0, 0) & (0, 0, 0)  & (0, 0, 0)  & (0, 0, -0.050) \\
$\bm{J}_3/M^2$ & (0, 0, 0)  & (0, 0, 0)  & (0, 0, 0) & (0, 0, 0)  & (0, 0, 0)  & (0, 0, -0.050) \\
$M_{\mathrm{ADM}}$ & 0.904 & 1.001 & 1.003 & 1.004 & 1.008 & 1.007 \\
\br
\end{tabular}
}
\end{table}

The set corresponds to the Einstein field equations in vacuum only if some topological constraints are satisfied. As discussed in \cite{Zlochower2005accurate, Akbarian2015black, alekseenko2004constraint, Gundlach2006well, brown2009covariant}, all $3+1$ fields have to obey the Hamiltonian and momentum constraints, hereafter written in terms of the conformal transformation:
\begin{eqnarray}
 \label{ham}
 && \mathcal{H}= R - \widetilde{A}_{lm} \widetilde{A}^{lm} + \frac{2}{3} K^2 = 0, \\
   \label{mom}
 && \mathcal{M}^i  = \partial_k \widetilde{A}^{ik} + \widetilde{\Gamma}^i_{lm} \widetilde{A}^{lm}- \frac{3}{2 \chi } \widetilde{A}^{ik} \partial_k \chi
   - \, \frac{2}{3} \widetilde{\gamma}^{ik} \partial_k K = 0.
\end{eqnarray}
During the numerical simulations, we also enforce $Tr \{\widetilde{A}_{ij}\}=0$, where $Tr\{\dots\}$ stands for the trace of a tensor, and impose the determinant of the conformal three-metric to take on unit value, that is $\widetilde{\gamma} = 1$. We point out that in all the evolution equations, for better numerical stability, wherever it is not differentiated $\widetilde{\Gamma}^i$ is replaced by $-\partial_j \widetilde{\gamma}^{ij}$ \cite{Campanelli2006accurate}.

\subsection{The \texttt{SFINGE} code}
Our algorithm is based on the use of spectral methods. In particular, the spatial derivatives of each field are computed in a Cartesian geometry by applying standard Fast Fourier Transforms (FFTs). For each time $t$, at each spatial collocation point ${\bm{x}}$, any field is decomposed as $f({\bm{x}}, t)= \sum_{{\bm{k}}} \widetilde{f}_{\bm{k}}(t)  \exp(i  {\bm{k}}\cdot{\bm{x}})$ on an equally spaced lattice of $N_x\times N_y \times N_z$ meshes. In the above decomposition, $\widetilde{f}_{\bm{k}}(t) \in \mathbb{Z}$ are the Fourier coefficients for each wavevector ${\bm{k}}= (k_x, k_y, k_z)$.
We consider the same physical length $L_0$ along each direction, therefore for each of them  the wavevector becomes $k = 2\pi m/L_0$ with a wavenumber $m=0, \pm 1, \dots \pm N_k$. Here, $N_k=N/2$ is the Nyquist mode and depends on the discretization. The code supports parallel architecture and makes massive use of MPI directives. As described in \textit{Meringolo et al.} \cite{Meringolo2021spectral}, the present numerical model has already been tested against several classical testbeds.

\begin{figure}
\centering
\includegraphics[width=0.98\textwidth]{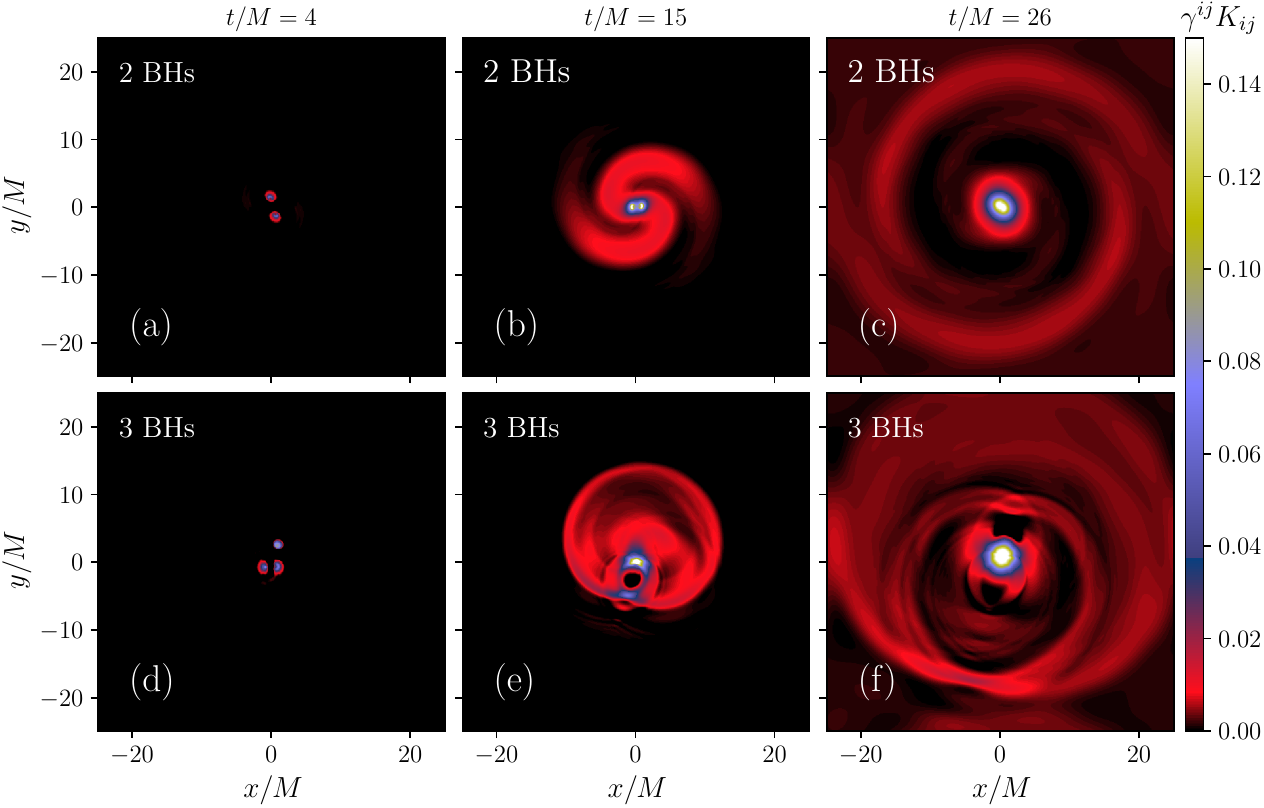}
\caption[ ]{\footnotesize{} Shaded contours of the extrinsic curvature $K$ for the two black holes (RUN I, top row) and the three-body `Classical Burrau' interaction (RUN II, bottom row), at three different times of the evolution, that is before, during, and after merger.}
\label{Figura2}
\end{figure}

The nonlinear terms in equations (\ref{evo1})--(\ref{slicing3}) can be evaluated through several transform-based techniques \cite{Canuto2012spectral, Boyle2007testing, Cooley1965algorithm}. Numerical artifacts may appear because of ``aliasing instabilities'' \cite{Hossain1992computing}, since any product among fields generates new modes with higher $k$'s, causing an alias when they become larger than the maximum available mode, namely at the wavenumber $N_k$. This problem, particularly recurrent in incompressible hydrodynamics, can be avoided by adopting dealiasing techniques \cite{Orszag1971elimination}. We then use a strategy to suppress such a problem based on analogous studies in compressible hydrodynamics, by using a filtering method \cite{Frisch2008hyperviscosity, Shu2005numerical, Meringolo2021aliasing, Meringolo2022pseudo}. For each decomposition in the spectral space, one can define 
\begin{eqnarray}
    f({\bm{x}}, t) = \sum _{{\bm{k}}} \widetilde{f}_{\bm{k}}(t) e^{i k_j x^j} \Phi_{k^*}({\bm{k}}),
\label{fltr}
\end{eqnarray} 
where $\Phi_{k^*}({\bm{k}})= e^{-a\left[\frac{|{\bm{k}}|}{k^*}\right]^a }$ represents the anti-aliasing filter, $a\,(=20)$ is a free parameter that regulates the smoothness of the filter at a certain cutoff $k^*$. Hereafter, we set $k^* = N/2.5$, where $N = 512$ is the number of grid points along each spatial direction.

To advance in time the BSSN fields, we implement a second-order Runge-Kutta scheme, with a variable (optimized) time-step \cite{Meringolo2021spectral}. Finally, we introduce an implicit, viscous technique capable of suppressing boundary disturbances, typical of numerical configurations that slightly violate the periodicity. In order to simulate spherical domains on a Cartesian geometry, we perform an implicit hyperviscous scheme with a radial envelope \cite{Schneider2005decaying, Dobler2006magnetic, Servidio2007compressible}. Terms of the type $\nu_n \nabla^n f$, where $\nu_n$ is a dissipative numerical coefficient, can indeed strongly dissipate very small-scale ripples and noise. See \cite{Meringolo2021spectral} and \cite{Meringolo2021aliasing} for more technical details on the procedure.

\section{Multiple black hole dynamics}
\label{sec:inspir}

The \texttt{SFINGE} code also includes modules to build initial data for a variety of problems, in a self-consistent way. 
In order to impose a set of well-posed initial conditions at some initial instant of time, we implement an iterative algorithm based on the same pseudo-spectral structure of the main code. Multiple black hole initial data have been constructed for configurations that anticipate EGI events in the classical case, but live in the final coalescing stage in the corresponding GR case. This is in line with the procedure illustrated by \textit{Campanelli et al.} \cite{Campanelli2006accurate}, where the black holes in the binary are relatively close at the beginning. Precisely, they are travelling on the Innermost Stable Circular Orbit (ISCO), which means immediately before the merging and ring-down phases. We use the method introduced by \textit{Brandt and Br{\"u}gmann} \cite{Brandt1997simple}, assuming a small violation of the requirement that the black holes are initially sufficiently far apart. In this sense, we verified that the iterative scheme converges and that the initial violation of the constraints is low enough.

\subsection{Initial data}
\label{ID}
The starting point for multiple black hole data without boost and spin is the Schwarzschild solution in isotropic coordinates. In the case of time symmetry, these conformally flat data can be generalized to an arbitrary number of black holes by adding the individual contribution in the conformal factor, that is $\psi = \chi^{-1/4}$. The initial conditions for most of our general relativistic simulations all contain non-zero velocities. In such cases, the momentum constraints do not automatically vanish as in the case of time-symmetric initial conditions. Therefore, the initial values for the $A_{ij}$ tensor contributions to the trace-free part of the extrinsic curvature will depend on both the linear and angular momentum of the black holes. We used an iterative Newton-based technique in order to impose such boosts and spin to the black holes.

Time-symmetric initial configuration of multiple black holes is known as Brill-Lindquist data \cite{Bowen1980time}, which can be considered good test cases for numerical codes, although their physical relevance is quite poor since we expect black holes to have linear momentum and spin. In this case, the momentum constraints do not vanish trivially and the conformal factor takes the following form \cite{Brandt1997simple}
\begin{eqnarray}
  \label{BL}
  \psi = 1 + \sum_k \frac{M_k}{2 r_k} + u \,\, = 1 + \psi_0 + u,
\end{eqnarray}
where $M_k$ and $r_k$ are respectively the mass parameter and the distance from the $k^{th}$ black hole from the origin of coordinates, and $u$ is a corrective term. To build the initial condition for multiple black holes with certain boost and spin, we follow the \textit{Bowen-York approach} solving the constraints of the momentum analytically and the Hamiltonian constraint numerically \cite{Alcubierre2008introduction, Cook2000initial, Baumgarte2010numerical}.

Firstly, we assume that the conformal metric is flat ($\widetilde{\gamma}_{i j} = \delta_{ij}$), the physical metric is asymptotically flat ($\lim_{r \to \infty} \psi = 1$), and the trace of the extrinsic curvature is identically zero (maximal slicing, $K = 0$), so that the momentum constraints reduce to a simple tensorial equation that can be solved analytically (see Chapter 3.2 of \textit{Baumgarte and Shapiro} \cite{Baumgarte2010numerical}). The resolution allows us to obtain an expression for the conformal trace-free part of the extrinsic curvature 
\begin{eqnarray}
   \nonumber
   \widetilde{A}_{ij} = \psi^{-4} A_{ij} = \sum_{k=1}^N \psi^{-6} \bar{A}_{ij}^{(k)},
   \\
   \nonumber
   \bar{A}_{ij}^{(k)} =  \frac{3}{2 (r^{(k)})^2} \Big[ P_i^{(k)} n_j^{(k)} + P_j^{(k)} n_i^{(k)} - \Big( \delta_{ij} - n_i^{(k)} n_j^{(k)} \Big) P^{l \, (k)} n_{l}^{(k)}  
   \\
   ~~~~~~~~~~~~~
   \label{extrcurv}
   + \, \frac{2}{r} \Big( \varepsilon_{ilm} S^{l \, (k)} n^{m \, (k)} n_j^{(k)} + \varepsilon_{jlm} S^{l \, (k)} n^{m \, (k)} n_i^{(k)} \Big)   \Big],
\end{eqnarray}
where the over-bar notation indicates a further conformal transformation, $n_i^{(k)} = x_i^{(k)} / \sqrt{\left(x^{(k)} \right)^2 + \left(y^{(k)} \right)^2 + \left(z^{(k)} \right)^2}$ is the unit normal vector pointing away from the $k^{th}$ black hole's center in the flat conformal space, $P^i=(P^x, P^y,P^z)$ and $S^i=(S^x, S^y,S^z)$ are constant vectors, standing for the linear and the angular momentum, respectively, and $\varepsilon_{ijl}$ is the completely antisymmetric Levi-Civita tensor in three dimensions.

Secondly, we proceed to solve numerically the Hamiltonian constraint which, recalling the form of the conformal factor in equation (\ref{BL}) and the assumptions we made, reduces to an elliptic equation for $u$ \cite{Alcubierre2008introduction}:
\begin{eqnarray}
    \label{perturbterm}
    D^2 u + \frac{1}{8 \psi_0^7} \bar{A}_{ij} \bar{A}^{ij} \left(1 + \frac{u}{\psi_0} \right)^{-7} = 0,
\end{eqnarray}
where we have enforced the fact that $K = 0$ and that the spatial metric is conformally flat, so that $\bar{R} = 0$. To solve such an elliptic equation, we have used an iterative Gauss-Seidel algorithm \cite{albu2002gauss}, starting from an initial guess, as specified in \textit{Cao et al.} \cite{cao2008reinvestigation}.

The initial data for multiple black holes contain more conditions and a longer convergence time for the iterative algorithm. In particular, these calculations might take several CPU hours by using the above Gauss-Seidel technique. To accelerate the convergence, we start on a low-resolution initial cube ($64^3$ mesh points), solving the Hamiltonian constraint relatively quickly. Then, to interpolate on a finer lattice (i.e. reducing the grid spacing), we first move on the Fourier space by increasing the number of $k-$vectors via zero-padding, and then we obtain the physical fields performing an inverse FFT. This allows us to have a better initial guess on the higher-resolution box, with a much faster converge rate. By iterating this technique, adapting recursively (by interpolation) the solution to finer grids, we finally get the initial data. As a remark, when one deals with multiple BHs, it is important to estimate their apparent horizons (AHs) \cite{cadez1974apparent, jaramillo2011study, atlas2022basics, pook2021what, lin2007new}. Our initial data for the 2-BHs are the same as in \textit{Campanelli et al.} \cite{Campanelli2006accurate}, where it has been shown that each of the BH is a well-separated body with distinct AH (and no common initial AH). For our 3-BHs cases, since we have comparable masses, linear momenta, and separations as in Campanelli's work, it is reasonable to expect similar AH configurations. In addition, we estimated the ellipsoidal AH shapes, typical of boosted (and spinning) black holes, as the initial guesses described in \textit{Huq et al.} \cite{huq2002locating}. The surfaces are safely separated at the beginning of the EGIs (not shown here).

\begin{figure}
\centering
\hspace{-10pt}  
\includegraphics[height=125mm,width=150mm]{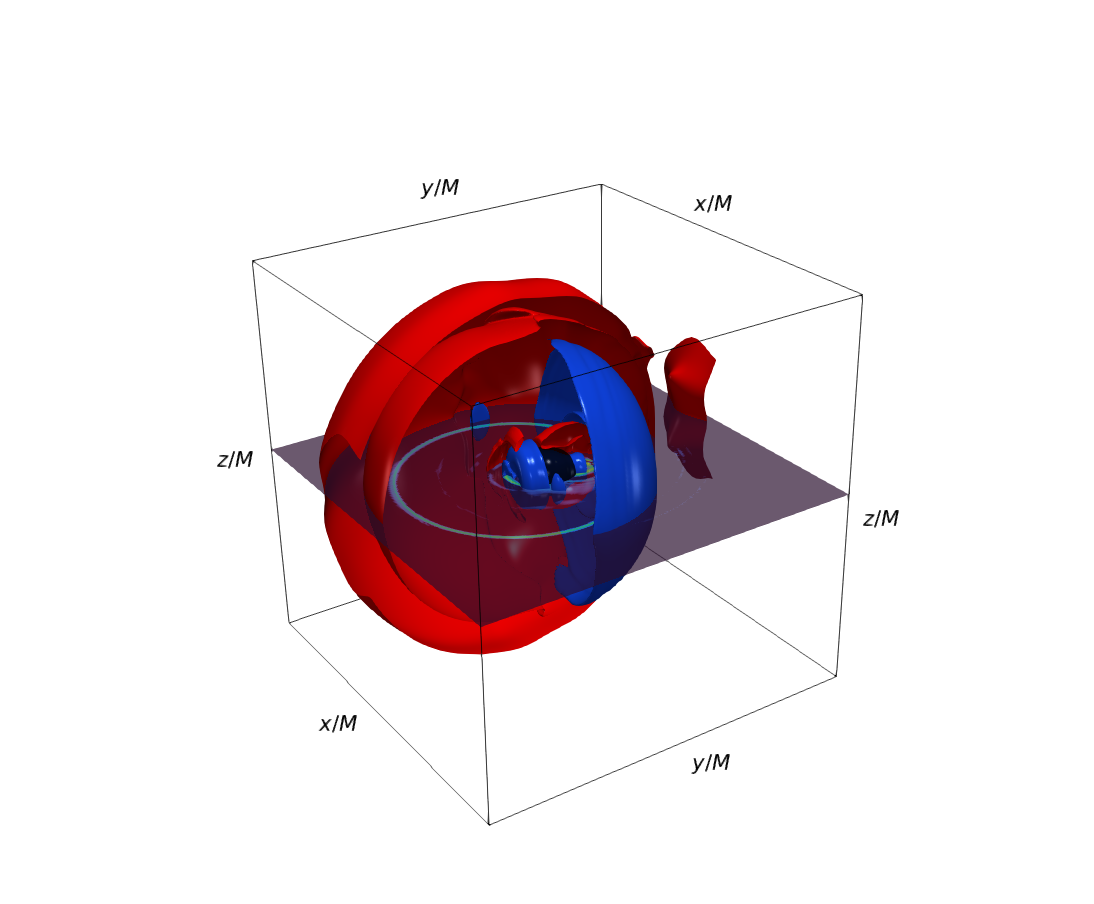}
\caption[ ]{\footnotesize{}3D representation of the three-black hole interaction, at time $t/M = 26$ for RUN II (``Classical Burrau'' in  Table \ref{TABLE2}). The (red and blue) isosurfaces represent constant values of the extrinsic curvature $K$, outside from the final black hole (central black-shaded region). The 2D transparent plane represents a color contour of the Newman-Penrose scalar $\Psi_4$, with maxima in light-green color.}
\label{Figura3}
\end{figure}

\subsection{Results}

Concerning the simulation parameters, for all the BSSN simulations, the domain is represented by a cube of side $L_0 = 50 M$, with $512^3$ mesh points. The time step is initially chosen to be $\Delta t = 8 \times 10^{-3}$. Furthermore, we check, for each case, that the Hamiltonian and the momentum constraints are well satisfied. The parameters of the BSSN simulations are summarized in Table \ref{TABLE2} and are labeled with roman numbers. Specifically, we are interested in the fully-relativistic evolution of the classical system when it is approaching an EGI, verifying that, at the initial time $t_0$, the BHs are sufficiently far apart. On the other hand, we avoid too large distances that would require an excessive numerical cost.

Before treating the complex three-body problem, we first calibrate our numerical experiments via the two-body inspiralling, which will provide us with the classical waveform \cite{Baker2001plunge, Baker2002modeling, baker2006binary}. The scope is to compare the properties of the gravitational waves generated from both the binary and the three-body systems. The numerical experiment describes the evolution of two compact objects of equal masses that are initially placed symmetrically with respect to the $x$-axis. They have a small initial velocity which is the same for both bodies but in opposite directions, and the properties of such ``2 BHs'' configuration are summarized in Table \ref{TABLE2} as RUN I. As expected, the BHs spiral around each other covering a single orbital period before merger, as shown in Fig.~\ref{Figura2}. A simple, fast way to identify a gravitational disturbance is to look at a scalar measure of the value of the extrinsic curvature, especially in the ecliptic plane. As can be seen from panels (a)--(c), small amplitude waves propagate out during the merger. After that, a large amplitude modulation propagates away, once the two marginally outermost trapped surfaces (MOTSs) come into contact \cite{lang2006measuring, centrella2010black, cotesta2018enriching, Brugmann2004numerical, sperhake2013numerical}. At this point, several disturbances of smaller amplitude begin to propagate until the system converges into a  Kerr-type black hole.

In the second campaign of GR simulations, we describe the evolution of 3-BHs, inspired by Run a--d. In particular, we take from these cases the initial data at $t_0$, as described in Fig.~\ref{Classical_cases}. With this strategy, by using spectral methods, we can concentrate on events that might be energetically relevant (and hence detectable by experiments). As reported in Table \ref{TABLE2}, RUN II represents the ``Classical Burrau'' case as in Run $a$, but with masses normalized such that the total mass of the system is set to unit value; RUN III (``Equal masses'') describes a similar system, but this time the masses of the black holes are equal, as in Run $b$; RUN IV describes a case similar to RUN II, but with different initial configuration (positions and momenta of BHs). All the above configurations refer to the case in which the three bodies are at rest at the very beginning, so we do not expect any global inspiralling. The ``Spinning Burrau'' (Run V) covers this case, namely the initial configuration is such that the angular momentum is different from zero (Run $d)$. We performed the last simulation, proper of GR, named ``Intrinsic spin'' (RUN VI), which presents the same initial setup as RUN V, except that the BHs have their own spin.

\begin{figure}
\centering
\includegraphics[height=125mm,width=150mm]{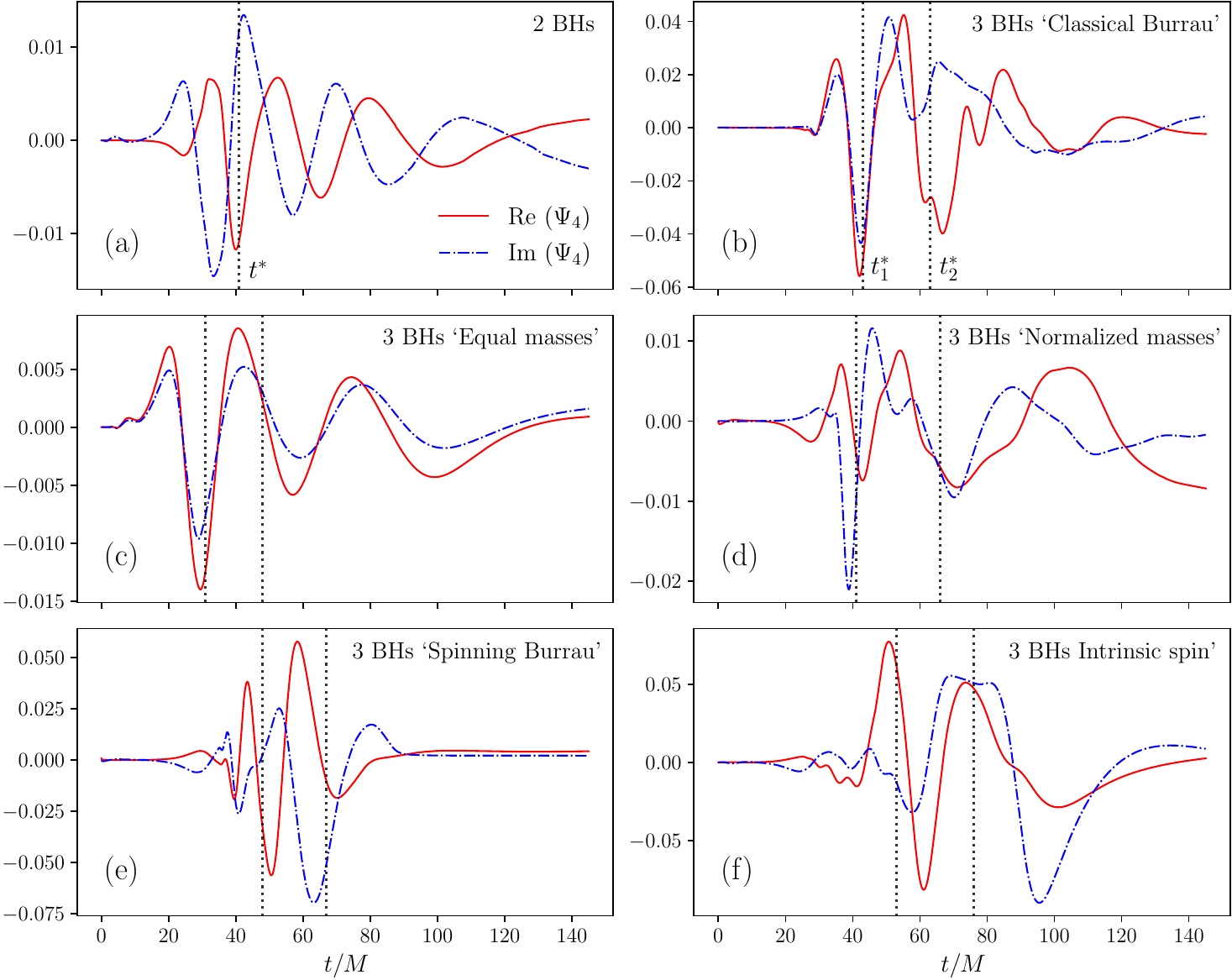}
\caption[ ]{\footnotesize{} Reconstruction of the gravitational wave signal for (a) the binary black hole system and (b)--(f) the three-black holes systems. Figures (a)--(f) refer respectively to RUNs I--VI listed in Table \ref{TABLE2}. The solid (dash-dotted) lines represent the real (imaginary) part of the spherical projection of $\Psi_4$; the vertical dotted black lines represent the merging time(s) of the black holes ($t^*$ is related to the 2-BHs merger, while $t^{*}_{1}$ and $t^{*}_{2}$ are related to the 3-BHs mergers), defined as in \cite{brugmann2008calibration, Buonanno2007inspiral, Baker2006binary}. All the simulations are performed up to $t/M = 145$. The three-black hole cases reveal a more structured and nonlinear behavior.}
\label{Figura4}
\end{figure}

Focusing on the overview of $K$ reported in Fig.~\ref{Figura2} (d)--(f), one can see how, compared to the 2-BHs, the patterns are more distorted, with finer scales. 
In the post-merging phase, after the system has been allowed to ``relax'', it is possible to see a different pattern for the extrinsic curvature, suggesting a different kind of asymptotic solution for the final black hole. In Fig.~\ref{Figura3}, we report a full-dimensional rendering of this RUN II, where we represent the gravitational radiation after the subsequent merging events. The (red and blue) shaded contours represent the isosurfaces of constant $K$, at two different values, showing how the outgoing waves are irregular and skewed. The region close to the final event horizon is represented with a (black) sphere.

To get more insight into the multi-black hole dynamics, we analyze the waveform by computing the projection of the Newman-Penrose scalar $\Psi_4$, which quantifies the radiation from a source of gravitational waves. This standard procedure represents the relativistic analog of the Poynting theorem in electromagnetism \cite{villalba2020newman, newman2009spin, campbell1977algebraic, iozzo2021extending, Alcubierre2008introduction, Baumgarte2010numerical}. After we computed this scalar by following the procedure described in \cite{reisswig2011gravitational, brugmann2008calibration}, we interpolated $\Psi_4$ from a Cartesian to a spherical grid, on which we calculated the outgoing radiation via the spin-weighted spherical harmonic $Y^{(-2)}_{2 2} (\theta, \varphi)$. This projection extracts the dominant contribution of the emitted waves, which comes from the quadrupole mode ($l = 2$, $m = 2$). In Fig.~\ref{Figura3}, we report the (color) contour of $\Psi_4$ in the equatorial $xy$-plane (at $z/M = 25 $),  for the three-body case (RUN II). The outgoing gravitational radiation is evident.

We compared all the signals coming from these multiple black hole interactions, by looking at the outgoing radiation as measured away from the sources. We evaluated the above projection at a given distance from the center of the box, namely at $r^\star = 20 M$, as a function of time, while the disturbances fly through the virtual detector. 
We also defined the merging time $t^*$ as the moment when a single connected isosurface of the lapse $\alpha = 0.3$ forms around two approaching black holes \cite{brugmann2008calibration, Buonanno2007inspiral, Baker2006binary}. All these times have been reported in Fig.~\ref{Figura4} employing vertical dotted black lines.
It is interesting to highlight that since we are interested in EGIs, we do not describe the long-time ``chirped'' signal before the merger. 

The signals are depicted in Fig.~\ref{Figura4}, revealing net differences among the cases.
First, contrary to the case of the binary [panel (a)], the real and imaginary parts of the waveform of some of the 3-BHs cases are almost in phase. Second, while for the 2-BHs only one peak can be observed (corresponding to the single merger), in the other cases multiple-scale disturbances are present, due to the subsequent collapses. Third, the signal from the three-body problem is much more irregular and unpredictable, revealing the presence of higher-order nonlinearities \cite{centrella2010black, anninos1995three, Zlochower2005accurate, Campanelli2006accurate, Lousto2008foundations, Campanelli2008close, silsbee2017lidov}. These more structured signals demand a deeper statistical investigation, performed as follows.

\begin{figure}
\centering
\hspace{-10pt}  
\includegraphics[height=75mm,width=100mm]{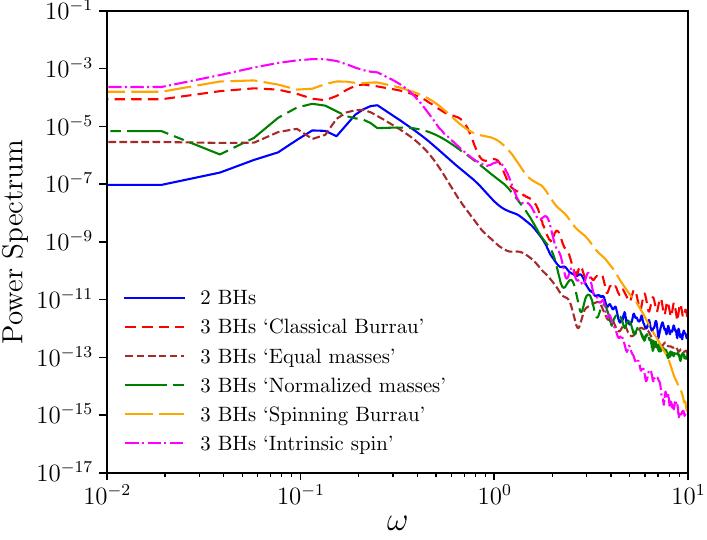}
\caption[ ]{\footnotesize{} Power spectrum of the waveforms in Fig.~\ref{Figura4}. The spectrum represents an average over the power spectra of the two signals (real and imaginary part of the projection of the Newman-Penrose scalar $\Psi_4$).}
\label{Figura5}
\end{figure}

To further highlight the differences among the configurations, identifying possible peculiarities of the 3-BHs interaction, we performed a spectral analysis of the metric disturbances shown in Fig.~\ref{Figura4}. As a first step, we smoothed the boundaries via a \textit{generalized normal window}, then we computed the Fourier transform and hence the power spectrum. We cross-checked the spectrum with an analogous procedure, by using the Blackman-Tukey technique \cite{blackman1958measurement}, transforming the windowed auto-correlation function of the signal. The spectrum has been computed for both the real and the imaginary parts of the projection appearing in Fig.~\ref{Figura4}. Finally, to decrease statistical uncertainties, we produced an average of the powers. As can be noticed from these averaged power spectra in Fig.~\ref{Figura5}, the 3-BHs cases produce in general a broader distribution of frequencies. The excess of high frequencies reflects the small scales features observed in the vicinity of the three-body system, already noticed in Fig.s \,\ref{Figura2} and \ref{Figura3}. The narrower spectrum is observed for the binary system and for RUN III, in which the head-on of the smallest black holes did not produce enough nonlinearity. On the other hand, the ``Spinning Burrau'' case produces the smallest scales of our campaign of simulations. It is worth noticing that in the three-body case the spectrum manifests a power-law behavior for $0.6<\omega<3$, stimulating interesting speculations about the possibility of a \textit{gravitational turbulent cascade}. However, the classical one-dimensional Fourier spectrum alone cannot say too much about the nature of the dynamics, since the processes are highly non-stationary. In this regard, we shall proceed with a refined analysis, described hereafter.

\begin{figure}
\centering
\includegraphics[height=105mm,width=160mm]{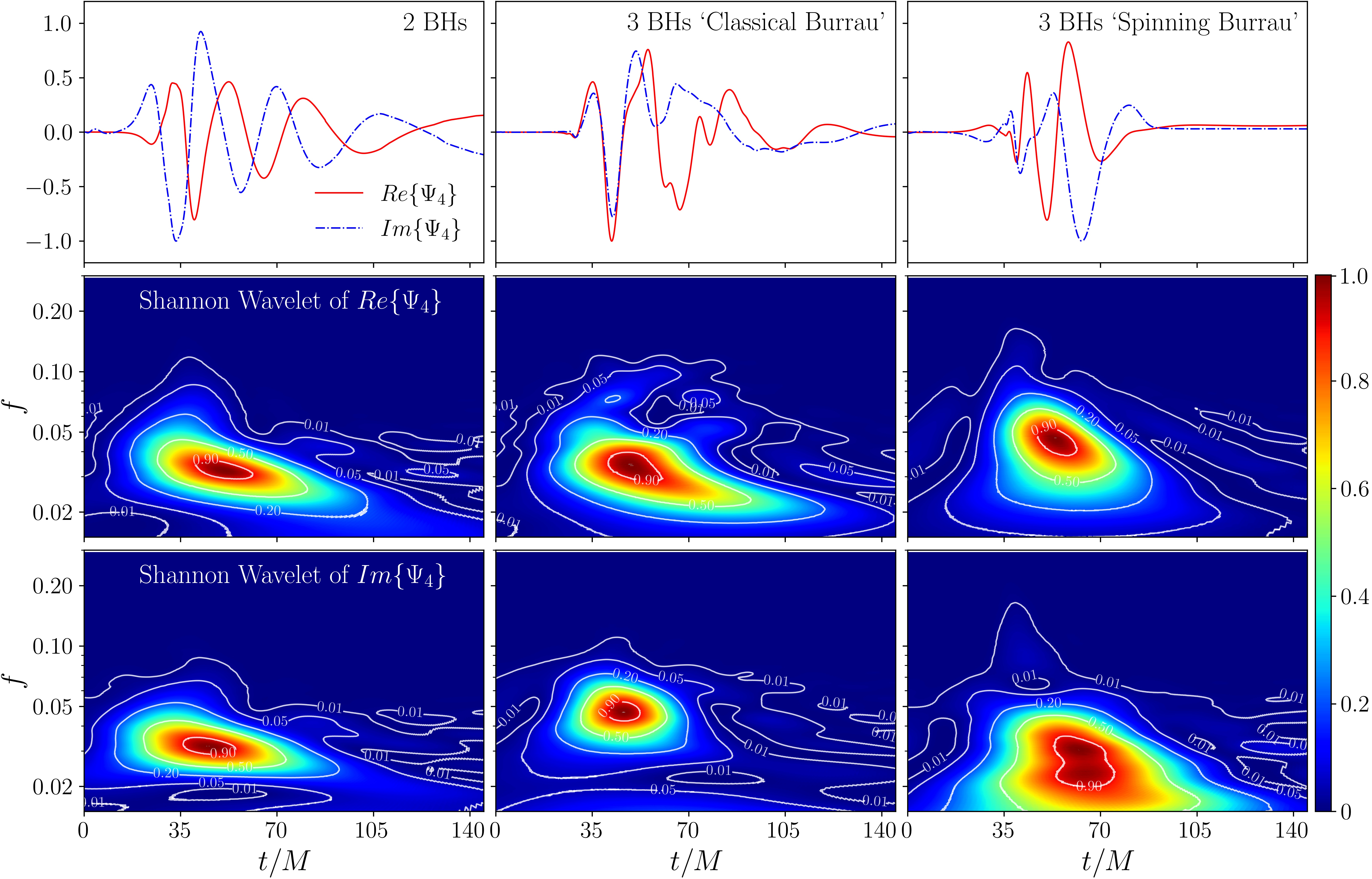}
\caption[ ]{\footnotesize{} Spectrogram of the Shannon wavelet (second and third row) for three different waveforms (first row), namely for RUN I (left), RUN II (center), and RUN V (right). In the middle and bottom row, we report the wavelet analysis for the real and the imaginary part of $\Psi_4$, respectively. The signals and the spectra have been normalized here to one to favor the comparison.}
\label{multiplot}
\end{figure}

Because of the non-stationarity of the signal, we analyze the wavefront by employing a wavelet decomposition. The main feature of wavelets is their natural splitting of fluctuations into different scale components according to the multiscale resolution analysis \cite{daubechies1992ten}, being therefore a powerful tool of investigation for multi-body coalescence events. Among the several choices, all of them qualitatively consistent with each other, the clearest results are here obtained for the Shannon generating function \cite{unser2000sampling}. This reconstruction makes use of the ``sinc function'', which is very localized and rapidly decaying to zero -- very adequate for the presence of discrete frequencies although  less performant for time-localization \cite{Cattani2008Shannon, mallat1999wavelet}. Such a typical wavelet spectrogram is reported in Fig.~\ref{multiplot}, where we take into consideration three characteristic cases: RUN I, II, and V. The 2-BHs case manifests a clear single peak in frequency, corresponding to the merging event.  A rise-up in frequency can be seen for $5<t<40$, typical of the inspiral phase \cite{Campanelli2006accurate, centrella2010black}. Real and imaginary signals of the scalar have very similar behavior. At a later time, the system shows the typical low-frequency modulation of the resultant, perturbed black hole. Very different is the behavior of the 3-BHs. In the Classical Burrau, by looking at $Re\{\Psi_4\}$, apart from the main frequency similar to the 2-BHs due to the first head-on merger, there are secondary peaks. There is a secondary peak at $t\sim 70$ and $f\sim 0.08$, due to the follow-up disturbance that propagates from the secondary EGI, when the larger BH {\it eats} the remaining one. This corresponds to a disturbance coming from large mass ratio coalescence. Finally, we performed the same analysis for the case with the broadest Fourier spectrum in Fig. \ref{Figura5}, namely for RUN V. In the Spinning Burrau, indeed, the BHs are confined by the large global angular momentum. This constraint forces the two mergers to happen at about the same time, with multiple nonlinear effects. This highly nonlinear interaction can be observed as a net broadening of the wavelet spectrum. In this case, the outgoing gravitational waves are carrying away from the final event horizon the reminiscence of such extreme interaction.

\section{Conclusions}
\label{sec:concl}

The three-body problem is a classical example of complex dynamics, where the three interacting masses can experience a chaotic behavior. We concentrated on two different approaches, namely (1) the Newtonian approach and (2) Einstein’s theory of general relativity. Regarding the first part, we summarized the classical results and developed a very accurate Lagrangian code that solves the trajectories of the bodies, starting from the Pythagorean configuration. Such configuration immediately tends toward a nonlinear regime. During the evolution, the system experiences quiet transients, as well as very bursty behaviors, typical of intermittency \cite{lichtenberg1992applied}. We solved the trajectories accurately and identified such very intense spikes, the EGIs. For future works, it would be very interesting to compare with models in which post-Newtonian corrections are kept into account \cite{boekholt2021relativistic}.

In the second (and more challenging) part, we built a gravitational model to study the dynamics of EGIs in terms of GR. By using a sequence of conformal transformations, we briefly introduced the BSSN equations. With equations in hand, we summarized the main properties of the \texttt{SFINGE} code, an algorithm based on a filtered pseudo-spectral scheme. As guidance for our GR experiments, we considered the problem of the binary, in the final stage of their coalescence, inside the ISCO. To measure the emitted gravitational waves, we dealt with the problem of wave extraction. By using a spherical interpolation, and following the Newman-Penrose formalism, we measured the outgoing gravitational waves, as classically done for compact objects simulations. After the two-body simulations, we finally considered the three black hole interaction. We extracted from the Newtonian simulations the initial data preceding a strong EGI, when the bodies are sufficiently far apart. We monitored several quantities, such as the metric tensor and the extrinsic curvature of the system, that suggest intense emission during the multiple-coalescence events. In analogy with the 2-BHs, we used the numerical extraction technique and reconstructed the emitted waveform, which is highly irregular and nonlinear. We compared finally the binary merger and the extreme three-body interaction, by computing the power spectra of the signals for all the cases. We found net differences, essentially at high frequencies, where the 3-BHs system exhibits a broader distribution of power. The spectrum seems consistent with a power-law distribution which might suggest that gravitation, similarly to hydrodynamics, is subject to a turbulent cascade.

To get more insight into the outcome of the multi-BH radiation, we performed a refined analysis of the signals. In particular, to detect the main differences among the cases, we made use of a Shannon wavelet. This analysis reveals net differences between the 2- and the 3-BHs cases. While in the former the projection evidences a single dominating frequency, in the 3-BHs the wavelets suggest that there are multiple frequencies produced during the evolution, due to the multiple EGIs. Finally, in the case in which there is a global angular momentum that confines the BHs, there is a net broadening of the wavelet power, typical of nonlinear phenomena.

While the 2-BHs inspiral has been largely investigated in literature \cite{Brugmann2004numerical, cao2008reinvestigation, campanelli2006last, tichy2006black, sperhake2015numerical, vaishnav2007matched, campanelli2006spinning, Tichy2009long, baker2006binary}, there is much less documentation about 3-BHs EGIs \cite{Lousto2008foundations, galaviz2010numerical} and, at the present, there is no observational evidence. These results could be of interest for future observational campaigns, since we expect that, with modern technologies, one can achieve better signal-to-noise sensibility, investigating whether such spectacular events might sometimes occur in the Universe. 
In future works, we will inspect the statistical properties of the gravitational fields in such nonlinear regimes, by performing larger, higher-resolution, and longer numerical experiments. Last but not least, we will inspect the development of new AHs that might form around a pair or a triplet of black holes when they come closer than critical distances \cite{jaramillo2011study}.
The loss of energy and momentum due to gravitational-wave emission and the lack of closed orbits even in the GR two-body problem could drastically enhance the differences between Newtonian and Einsteinian gravity. In this regard, post-Newtonian approximations might allow comparison with semi-analytic cases \cite{Valtonen1995Burrau}.

\ack

Authors would like to acknowledge Jér\^{o}me Novak for his time and constructive discussions. The work has been supported by “Progetto STAR 2-PIR01\_00008” (Italian Ministry of University and Research). The simulations have been performed by using a parallel architecture based on MPI directives, at the Newton HPPC Computing Facility at the University of Calabria.

\appendix

\section{}
\label{appendix}

The constraints of the ADM formalism have to be satisfied for the entire duration of the simulation (see Section \ref{ID} for more details). For this reason, it may be useful to monitor how the Hamiltonian and momentum constraints are violated during the dynamics. 
\begin{figure}[t]
\centering
\includegraphics[height=100mm,width=135mm]{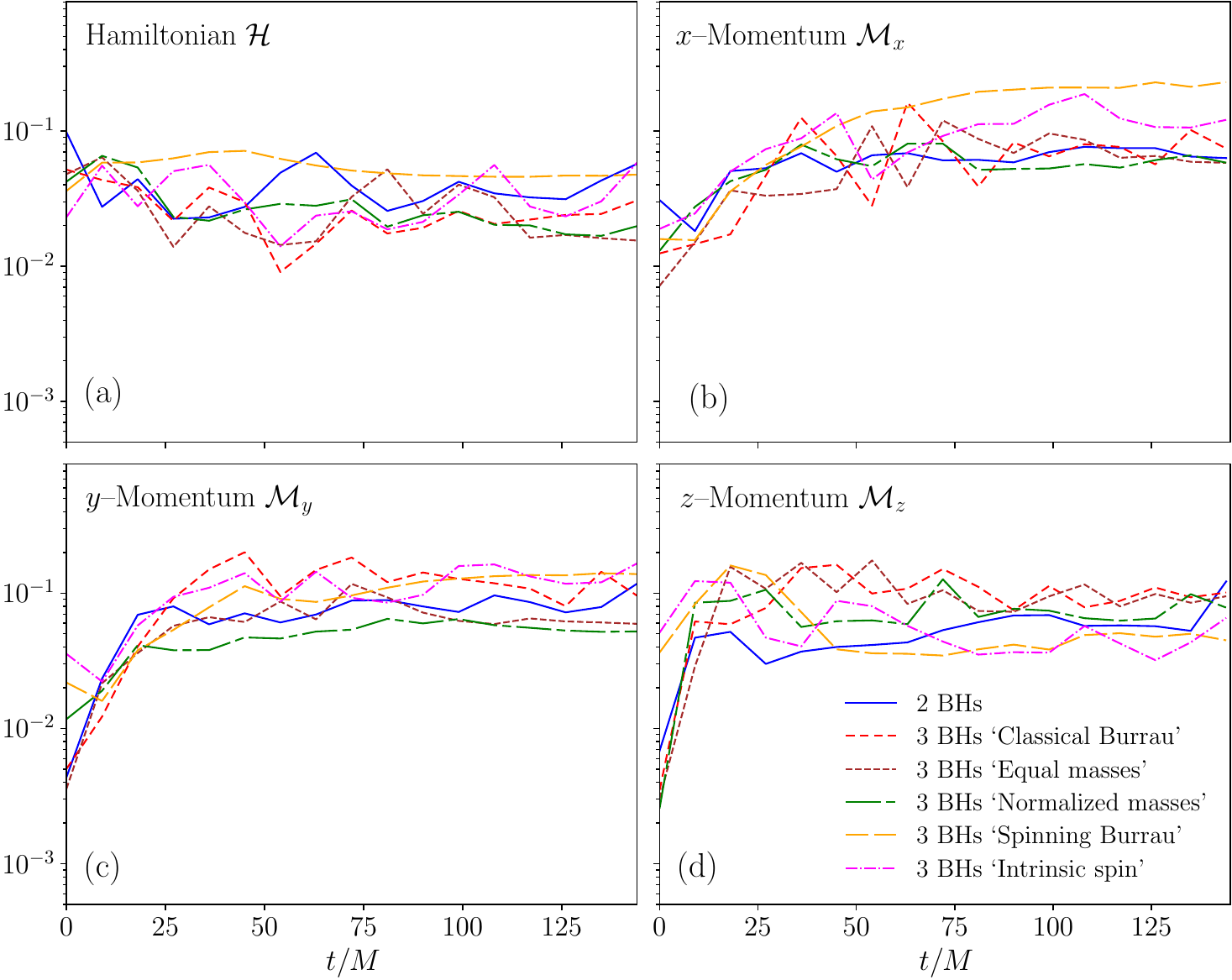}
\caption[ ]{\footnotesize{} Time evolution of the Hamiltonian constraint (a) and the Momentum constraints (b)--(d). The values on the $y$--axis are reported in logarithmic scale.}
\label{Figura8}
\end{figure}
In principle, one expects that the numerical solution satisfies them or at least leads to residuals that will not grow once the evolution of the resulting gravitational field is computed. The violation of the constraints is evaluated through their average $\bar{L}_2$ error, defined as \cite{Dumbser2018conformal}
\begin{equation}
    \label{L2}
    \bar{L}_2 = || \epsilon (x, y, z) ||_2 = \sqrt{\frac{\int_{\Omega} \epsilon^2 \sqrt{|\gamma|} d \Omega}{\int_{\Omega} \sqrt{|\gamma|} d \Omega}}.
\end{equation}
Here $\sqrt{|\gamma|} d \Omega$ represents the volume element and $\epsilon$ denotes the local error of each of the quantities we are interested in, namely the Hamiltonian $\mathcal{H}$ and the momentum constraints $\mathcal{M}_i$. 
We emphasize that the volume integrals in equation (\ref{L2}), in our case, have been computed over the whole computational domain, taking into account even the grid regions closest to the singularities. 
Fig.~\ref{Figura8} shows the time--dependent Hamiltonian and momenta for each of the analyzed configurations, whose trend expresses the violation of each of the constraints that must be satisfied at every time slice. As can be noted, the violation remains low and fairly constant through the evolution of the simulation, without any exponential growth.

\end{document}